\documentclass[a4paper]{jpconf}
\usepackage{graphicx,amsmath,amssymb}

\newcommand{\EPJ}{{\it Eur. Phys. J.} }
\newcommand{\PPNP}{{\it Prog. Part. Nucl. Phys.} }
\newcommand{\PREP}{{\it Preprint\/} }
\begin{document}

%%%%%%%%%%%%%%%%%%%%%%%%%%%%%%%%
%	Title
%%%%%%%%%%%%%%%%%%%%%%%%%%%%%%%%
\title{Amplitude analysis of $\bar{K}N$ scattering}

%%%%%%%%%%%%%%%%%%%%%%%%%%%%%%%%
%	Authors
%%%%%%%%%%%%%%%%%%%%%%%%%%%%%%%%
\author{C\'esar Fern\'andez-Ram\'{\i}rez}
\address{Instituto de Ciencias Nucleares, Universidad Nacional Aut\'onoma de M\'exico, 
A.P.~70-543, M\'exico D.F.~04510, M\'exico}
\ead{cesar.fernandez@nucleares.unam.mx}

%%%%%%%%%%%%%%%%%%%%%%%%%%%%%%%%
%	Abstract
%%%%%%%%%%%%%%%%%%%%%%%%%%%%%%%%
\begin{abstract}
We present the results of a coupled-channel model for $\bar{K}N$ scattering in the resonance region.
The model fulfills unitarity, has the correct analytical properties for the amplitudes 
and the partial waves have the right threshold behavior. 
The parameters of the model have been established by fitting single-energy partial waves up to $J=7/2$ 
and up to 2.15 GeV of center-of-mass energy. 
The $\Lambda^*$ and $\Sigma^*$ spectra has been obtained, providing
a comprehensive picture of the $S=-1$ hyperon spectrum.
We use the structure of the hyperon spectrum and Regge phenomenology 
to gain insight on the nature of the $\Lambda(1405)$ resonances.
\end{abstract}

%%%%%%%%%%%%%%%%%%%%%%%%%%%%%%%%
%	Introduction
%%%%%%%%%%%%%%%%%%%%%%%%%%%%%%%%
\section{Introduction}
The comprehensive understanding of the strong interaction 
in the hadronic energy range is one of the open problems in particle and nuclear physics.
Problems such as non-perturbative aspects of Quantum Chromo Dynamics (QCD),
how quarks and gluons aggregate to constitute hadrons, 
confinement and chiral symmetry breaking
have spectroscopy of excited baryons constituted 
by light valence quarks ($u$, $d$, and $s$) as one of their primary tools for research.
For these reasons,
many experiments have measured $\pi N$ and $\bar{K}N$ scattering
as well as photoproduction off the proton in order to garner information on the spectrum.
However, the amount of experimental information to pin down
hyperon resonances with $S=-1$ ($Y^*=\Lambda^*, \Sigma^*$) 
is not as wealthy as for the case of nucleon excitations ($S=0$)
and, as a consequence, its spectrum is less understood.  
Only in recent years we are starting to have a good grasp of the $Y^*$ spectrum and pole positions 
have started to be reported by the {\it Review of Particle Physics} (RPP) \cite{PDG2014}
thanks to the development of models for  kaon electroproduction \cite{Qiang10} and
$\bar{K}N$ scattering \cite{FR2015,KNmodels}.
Once a baryon resonance and its location in the complex plane have been well-established,
it is time to understand if the nature of that state is (mostly) that of a three quark system, a molecule or a pentaquark,
and to find out what is the role of gluonic excitations for that particular state.
In section \ref{sec:kn} we provide a brief description of the amplitude analysis of $\bar{K}N$ scattering performed 
in \cite{FR2015} that allowed to provide the most comprehensive picture of the $S=-1$ hyperon spectrum to the date. 
In section \ref{sec:lambda1405} we use the structure of the hyperon spectrum and Regge phenomenology 
to gain insight on the nature of the $\Lambda(1405)$ resonance, the first excitation of the isospin-0 $uds$ system
and a long-standing problem in hadron spectroscopy.

%%%%%%%%%%%%%%%%%%%%%%%%%%%%%%%%
%	KN model
%%%%%%%%%%%%%%%%%%%%%%%%%%%%%%%%
\section{$\bar{K}N$ scattering in the resonance region} \label{sec:kn}
The $\bar{K}N \to \bar{K}N$ reaction, besides its importance for studies of the $Y^*$ spectrum, 
plays a role in amplitude analysis of more complicated reactions, which 
include decays (pentaquark searches) \cite{LHCbpentaquark,INTworkshop} 
and  $K \bar{K}$ pair photoproduction \cite{ATHOS,JLAB}.
For example, the recent observation of two pentaquark states in  
$\Lambda_b^0 \to J \slash \psi \: K^- p$ decay \cite{LHCbpentaquark}  uses 
a specific model to incorporate  $Y^*$ resonances in the $K^- p$ channel.
Studies of systematic uncertainties should involve the comparison with other models of 
 $\bar{K}N$ interactions and should incorporate background effects, not just $Y^*$ excitations \cite{INTworkshop}. 
Real and quasi-real diffractive photoproduction of $K \bar{K}$ pairs can produce 
the poorly known \cite{ATHOS}, {\it i.e.}, mesons containing $s\bar{s}$ pairs that also include
exotic mesons with hidden strangeness. 
The factorization of the $K\bar{K}$ photoproduction vertex   
requires separation of target fragmentation at the amplitude level.    
Hence,  the provision of amplitudes describing the $\bar K N$ interactions in  target fragmentation 
is relevant to  future  partial-wave analyses of the $\gamma p \to K \bar{K} p$ process.   
At Jefferson Lab \cite{JLAB}, both CLAS12 (Hall B) and GluEx (Hall D) experiments will devote
part of their effort to study this reaction. Hence, it is timely to develop a model for $\bar{K}N \to \bar{K}N$ 
scattering that can be incorporated in the analysis of both three-body decay and two kaon photoproduction experiments.

In  \cite{FR2015} we presented a coupled-channel model for $\bar{K}N \to \bar{K}N$ scattering in the resonance region
that incorporates up to 13 channels per partial wave, analyticity, unitarity, 
and the right angular momentum barrier for the partial waves.
All the details can be found in \cite{FR2015} and here we sketch the building blocks of the model.

The partial-wave expanded $S_\ell$ matrix is related to the amplitude $T_\ell$ through
\begin{equation}
S_\ell=\mathbb{I}+2iR_\ell(s)=\mathbb{I}+2i \left[C_\ell (s) \right]^{1/2} T_\ell(s) \left[C_\ell (s) \right]^{1/2}, \label{eq:smatrix}
\end{equation}
where $\mathbb{I}$ is the identity matrix and the diagonal matrix
\begin{equation}
C_\ell (s) = \frac{q_k (s)}{q_0}\left[ \frac{r^2q^2_k(s)}{1+ r^2q^2_k(s) }\right]^{\ell},  \label{C}
\end{equation}
accounts for the phase space, where $q_k(s)  = \sqrt{( m_1 m_2) (s-s_k )}/ (m_1+m_2)$,
$s_k$ is the threshold center-of-mass energy squared of the corresponding channel $k$,
$m_1$ and $m_2$ are the masses of the final states, 
$q_0=2$ GeV is a normalization factor  for the momentum in the resonance region,
$r=1$ fm is an effective interaction range parameter
and $T_\ell(s)$ is the analytical partial-wave amplitude matrix. 
We write $T_\ell(s)$ in terms of a $K$ matrix  \cite{kmatrix} to ensure 
unitarity 
\begin{equation}
T_\ell (s)= \left[ K(s)^{-1} -i \rho(s,\ell) \:  \right]^{-1}, \label{eq:kmatrix}
\end{equation}
where $\rho(s,\ell )$  is the dispersive integral over the 
phase space matrix $C_\ell (s)$,  
\begin{equation}
i \rho (s,\ell)  =   \frac{s-s_k}{\pi} \int_{s_k}^\infty\frac{ C_\ell (s')  }{s'-s} \frac{ds'}{s'-s_k}. \label{eq:rho}
\end{equation}

The $K(s)$ matrix in equation (\ref{eq:kmatrix}) is built as the addition of up to six $K$ matrices
\begin{equation}
\left[ K(s) \right]_{kj} =  \sum_a x^a_k\:K_a(s)\: x^a_j  \:.
\end{equation}

Each one of the $K_a(s)$ matrices can be either a {\it pole} $K$-matrix
\begin{equation}
\left[ K_P (s)\right]_{kj} = x^P_k \: \frac{M_P}{M_P^2-s} \:  x^P_j \:, \label{eq:kr}
\end{equation}
or a {\it background} $K$ matrix
\begin{equation}
\left[ K_B (s)\right]_{kj} = x^B_k \: \frac{M_B}{M_B^2+s} \:  x^B_j \:, \label{eq:kb}
\end{equation}
depending on the considered partial wave. 
The free parameters $x^P_j$, $x^B_j$, $M_B$ and $M_P$ will be fixed 
by fitting single-energy partial waves from \cite{Manley13a}.

\begin{figure}[h]
\begin{minipage}{18pc}
\includegraphics[width=18pc]{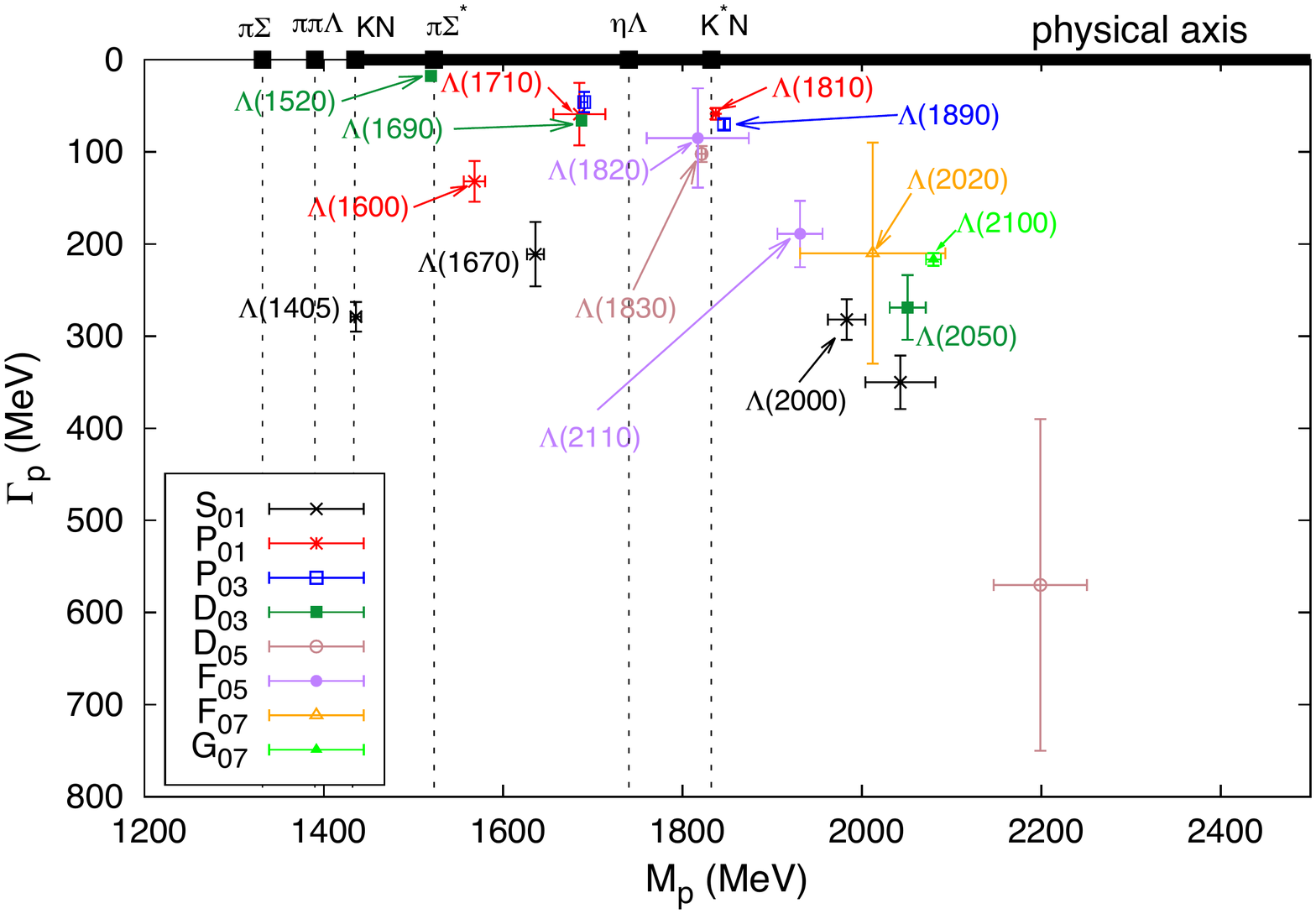}
\caption{\label{fig:spectrum0}Isospin 0 ($\Lambda^*$) poles. 
Vertical dashed lines mark the different production thresholds.}
\end{minipage}\hspace{2pc}%
\begin{minipage}{18pc}
\includegraphics[width=18pc]{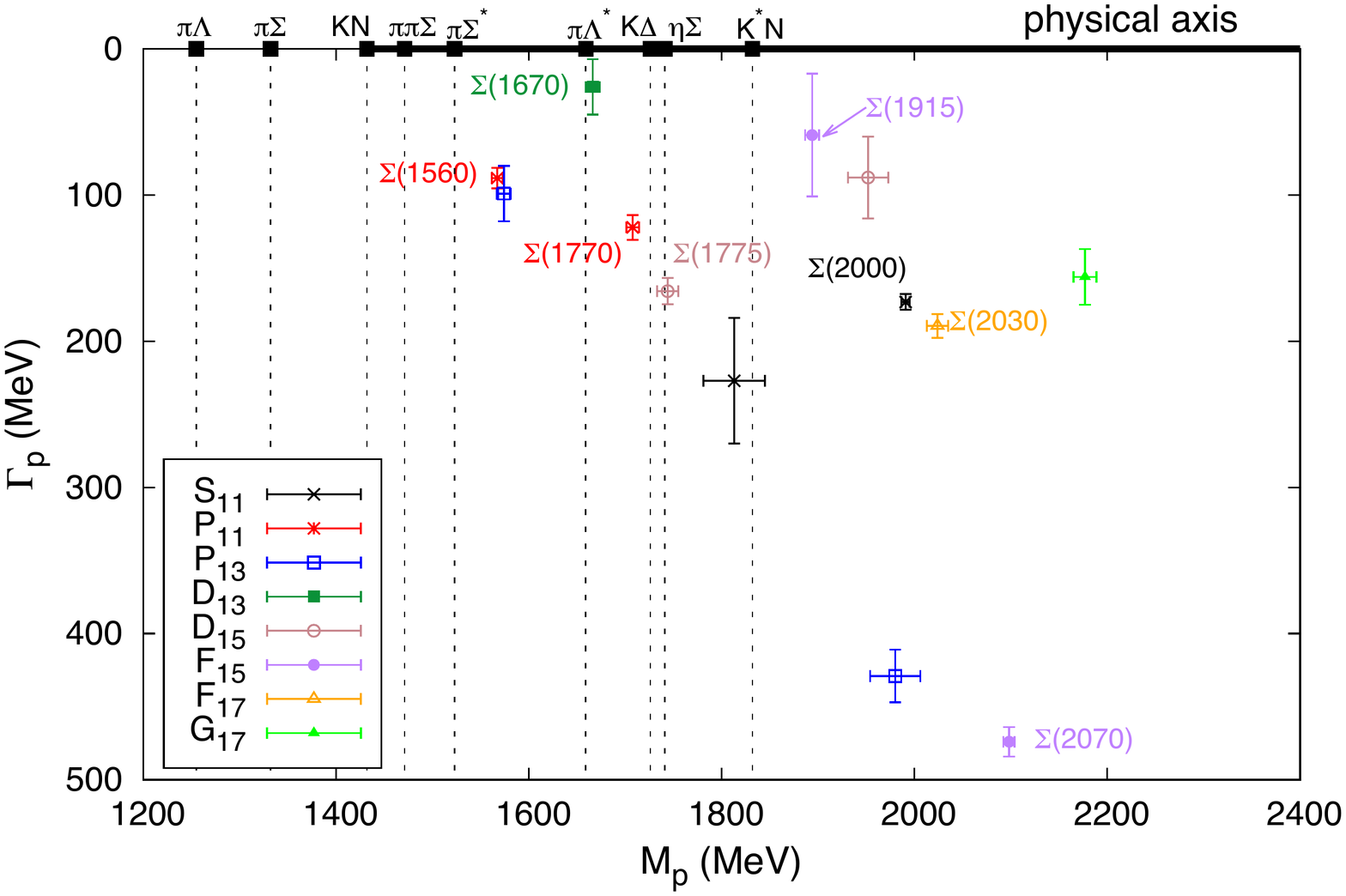}
\caption{\label{fig:spectrum1} Isospin 1 ($\Sigma^*$) poles. 
Vertical dashed lines mark the different production thresholds.}
\end{minipage} 
\end{figure}

In \cite{Manley13a} the experimental database in the resonance region with $2.19 < s < 4.70~ \mbox{GeV}^2$
(approximately 17500 data points for the 
$K^- p \to K^- p$, $K^- p \to \bar{K}^0 n$, $K^- p \to \pi^0 \Lambda$, 
$K^- p \to \pi^0 \Sigma^0$, $K^- p \to \pi^- \Sigma^+$, and $K^- p \to \pi^+ \Sigma^-$ channels)
was analyzed and the single-energy partial waves were obtained  for ($\ell_{I\: 2J}$) up to $J=7/2$,
{\it i.e.} $S_{01}$, $P_{01}$, $P_{03}$, $D_{03}$, $D_{05}$, $F_{05}$, $F_{07}$, 
$G_{07}$, $S_{11}$, $P_{11}$, $P_{13}$, $D_{13}$, $D_{15}$, $F_{15}$, $F_{17}$, and $G_{17}$.  
We fitted our model to these single-energy partial waves employing  \textsc{MINUIT} \cite{MINUIT} 
and a genetic algorithm \cite{genetic}. 
The uncertainties were computed using the bootstrap technique \cite{NumericalRecipes}.
The value of all parameters and the codes to compute the partial waves and the observables are available from the 
Joint Physics Analysis Center webpage  \cite{Mathieu16,JPACWebpage}.
Once the parameters of the model have been established we can analytically continue the partial waves to the unphysical 
Riemann sheets and search for poles that correspond to the hyperon resonances.
Figures \ref{fig:spectrum0} and \ref{fig:spectrum1} show the $\Lambda^*$ and $\Sigma^*$ 
resonances found in the different partial waves.

%%%%%%%%%%%%%%%%%%%%%%%%%%%%%%%%
%	Figure: Spectra
%%%%%%%%%%%%%%%%%%%%%%%%%%%%%%%%

%%%%%%%%%%%%%%%%%%%%%%%%%%%%%%%%
%	Lambda(1405)
%%%%%%%%%%%%%%%%%%%%%%%%%%%%%%%%
\section{The nature of the $\Lambda(1405)$ resonance(s)} \label{sec:lambda1405}
Since its discovery in the early 60's \cite{lambda},
the nature of the $\Lambda (1405)$, the first excitation of the isospin-0 $uds$ system, 
is a long-standing problem in hadron spectroscopy.
It has regained interest due to the recent experimental confirmation of its, for decades 
assumed, spin and parity $J^P=  \frac{1}{2}^-$ \cite{Moriya2014}
and the latest developments in amplitude analysis \cite{Mai2015,twopoles},
lattice QCD \cite{lattice,Hall2015,Engel13}, 
and quark-diquark  models \cite{Santopinto14,Faustov15}.

\subsection{Status of amplitude analysis in the $\Lambda(1405)$ region}
A recent combined amplitude analysis on $\bar{K}N$ scattering and
$\pi \Sigma K^+$ photoproduction off the proton data based on the
unitarization of the chiral potential via a coupled channel Bethe--Salpeter 
equation  \cite{Mai2015} has confirmed previous works 
\cite{twopoles,molecule} 
that stated that the $\Lambda (1405)$ 
compiled in the RPP \cite{PDG2014}
is not a single state 
but two different resonances (poles)
located at $1429^{+8}_{-7}-i \: 12 ^{+2}_{-3}$ MeV  and at
$1325^{+15}_{-15}-i\: 90 ^{+12}_{-18}$ MeV \cite{Mai2015}.
In \cite{Mai2015}  and previous analyses \cite{molecule}, 
the resonances are dynamically generated and their authors interpreted both poles as molecular states
whose origin is entirely due to the $\bar{K}N$ and $\pi \Sigma$ interactions.
However, an interpretation of this kind has to be taken with care
because, besides model-dependency, the employed chiral Lagrangian in \cite{Mai2015}
incorporates fourteen low-energy constants that were treated as free parameters 
with the only requirement of being of natural size. These constants encapsulate the
information on unresolved degrees of freedom and can hide resonant states 
whose origin is not dynamical.

\subsection{Status of $\Lambda(1405)$ in quark models}
The $\Lambda (1405)$ quantum numbers are those of a $uds$ state but, in the past, 
constituent quark models have failed to match the experimental mass \cite{quarkmodels}.
This situation has improved during 2015 with two calculations
based on the relativistic interacting quark-diquark model
that obtain values closer to the amplitude extractions:
1431 MeV in \cite{Santopinto14} and 1406 MeV in \cite{Faustov15}.

\subsection{Status of $\Lambda(1405)$  in lattice QCD}
Although it would be desirable to understand the nature of the $\Lambda (1405)$ 
directly from QCD via lattice QCD, the resonant nature of this state makes its numerical investigation taxing. 
There have been recent exploratory studies of this state from lattice QCD 
identifying the $\Lambda (1405)$ as a $\bar{K}N$ molecule in \cite{Hall2015} 
and as an ordinary three-quark state in \cite{Engel13}.
Unfortunately, in accessing the spectrum and the subsequent analysis of the spectrum the strong 
decay to the $\pi \Sigma,\bar{K}N$ scattering states is neglected. 
This approximation has been formally \cite{formally} 
and numerically \cite{Dudek:2012xn, mesonsector} demonstrated to only be justifiable 
in the narrow-width limit. 
In fact, it has been recently demonstrated that finite volume 
matrix elements of electromagnetic current obtained via lattice QCD, 
cannot be directly interpreted as infinite-volume form factors of resonances \cite{Briceno2015}
Hence, it is hard to asses the size of the systematic errors of the exploratory investigations listed above. 
In order to make a definitive statement of the implications of lattice QCD regarding 
the true nature of the $\Lambda (1405)$, a thorough analysis of the coupled 
$\pi \Sigma,\bar{K}N$  scattering amplitude will be needed, as has been done in 
\cite{mesonsector} for the mesonic sector.

\subsection{Regge analysis of the $\Lambda^*$ spectrum}
\begin{figure}[h]
\begin{minipage}{18pc}
\includegraphics[width=18pc]{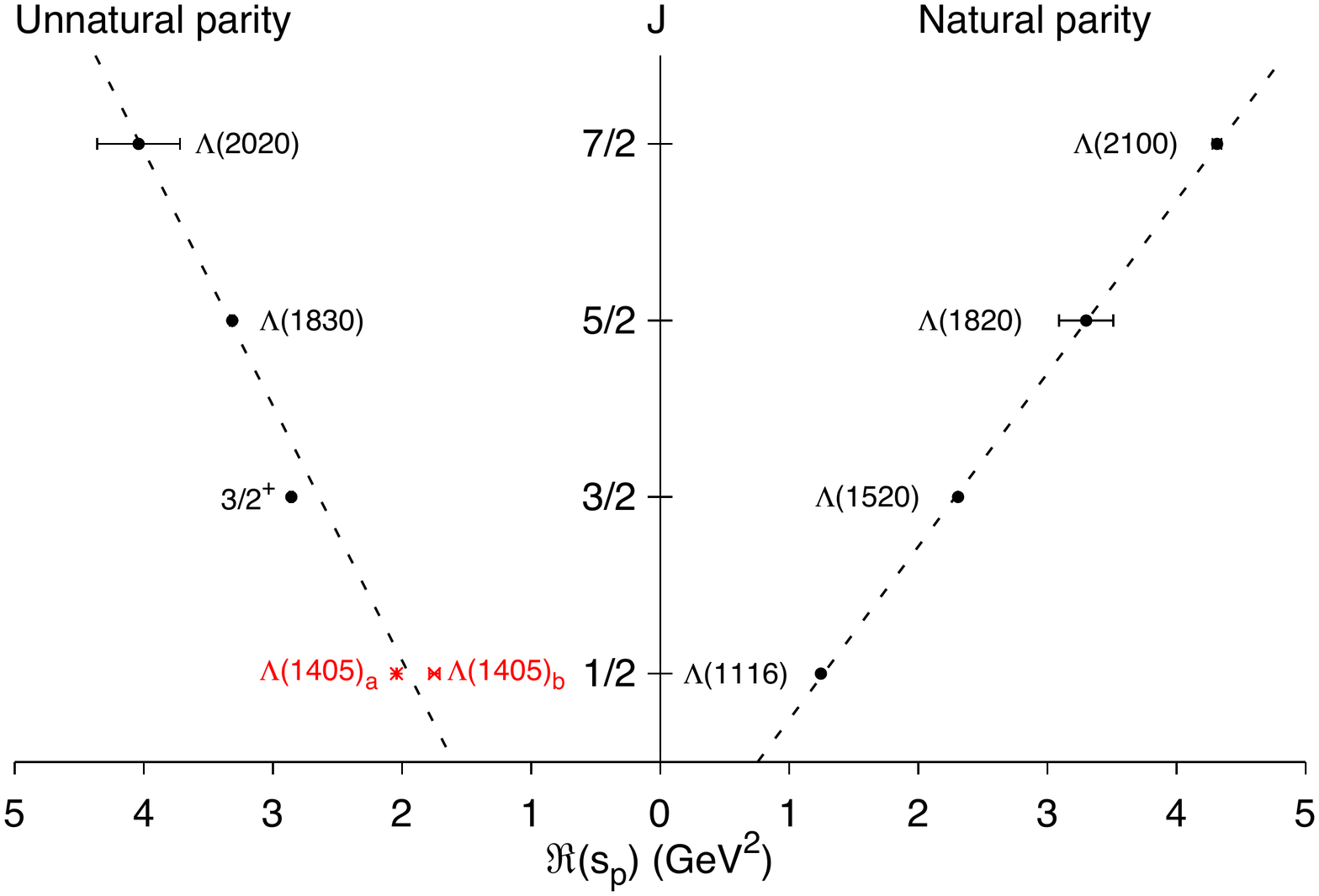}
\caption{\label{fig:revsJ} $\Re(s_p)$ {\it vs.} $J$ (Chew--Frautschi)
plot for the $\Lambda$ parent Regge trajectories. The dashed lines are to guide the eye.}
\end{minipage}\hspace{2pc}%
\begin{minipage}{18pc}
\includegraphics[width=18pc]{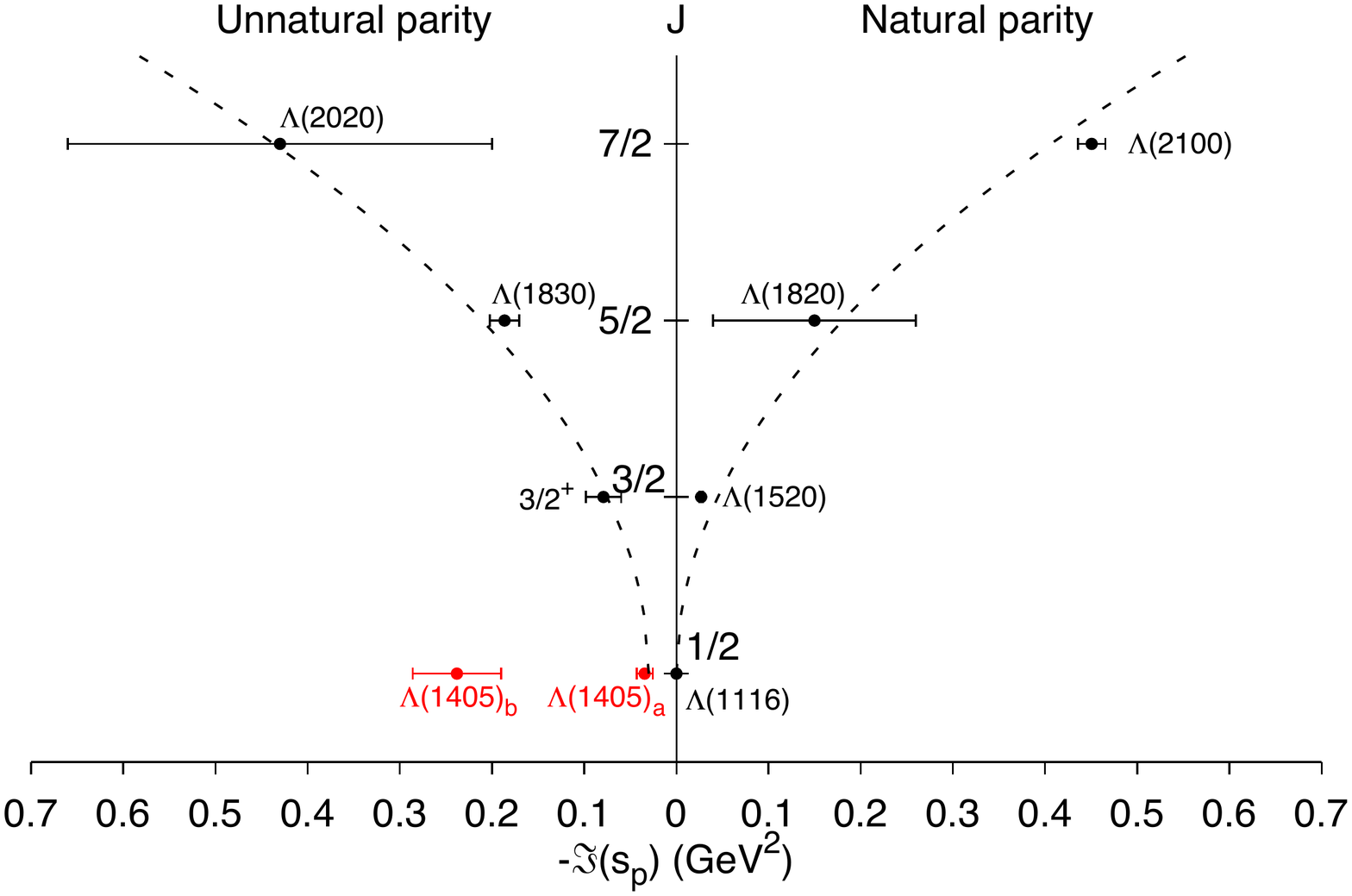}
\caption{\label{fig:imvsJ}$-\Im (s_p)$ {\it vs.} $J$ plot for the $\Lambda$ parent Regge trajectories. 
The dashed lines are to guide the eye.}
\end{minipage} 
\end{figure}

It is clear that a big progress has been made in the last years,
but the final answer on the nature of the $\Lambda (1405)$ state(s) remains elusive.
In \cite{JPACLambda1405} we used Regge phenomenology and our knowledge on the hyperon spectrum to 
assess the nature of the two $\Lambda (1405)$ poles.
From complex angular momentum theory
we know that resonances follow Regge trajectories that
connect poles at different partial waves \cite{Gribov}. 
Hence, we can use Regge trajectories to connect 
the ground states to their excitations and derive conclusions 
based on the shape of the trajectories, as their (non-)linearity depend 
on the quark-gluon dynamics \cite{quarklinear}.
If we use the current knowledge on the $\Lambda^*$
spectra we can compute the parent Regge trajectories and shed light on the nature of the $\Lambda (1405)$ states
by testing how well do the two poles fit within said trajectories.
In figure \ref{fig:revsJ} we show the Chew--Frautschi plot \cite{CFplot}, {\it i.e.} $\Re(s_p)$ {\it vs.} $J$,
where $s_p$ is the pole position and $J$ stands for the total angular momentum of the resonance.
The $\Lambda (1116)$ is taken from the RPP \cite{PDG2014},
$\Lambda (1405)_a$ ($1429^{+8}_{-7}-i \: 12 ^{+2}_{-3}$ MeV) 
and $\Lambda (1405)_b$ ($1325^{+15}_{-15}-i\: 90 ^{+12}_{-18}$ MeV) from \cite{Mai2015},
and the rest of the $\Lambda^*$ poles from the
$\bar{K}N$ amplitude analysis presented in section \ref{sec:kn}.
It is apparent how the projection of the Regge trajectories onto the ($\Re(s_p)$,$J$) plane in figure \ref{fig:revsJ}
follow linear trajectories as expected from three-quark states \cite{quarklinear}.
However, both $\Lambda (1405)$ resonances lay on top of the parent Regge trajectory.
The situation clarifies if we realize that the poles are complex quantities 
and it is also possible to plot $-\Im (s_p)$ {\it vs.} $J$, which corresponds to the 
projection of the Regge trajectory onto the ($-\Im(s_p)$,$J$) plane.
We provide such plot in figure \ref{fig:imvsJ}. 
It is apparent how the poles follow a square-root behavior and that the $\Lambda(1405)_a$ state fits in the 
unnatural parity parent Regge trajectory while the $\Lambda(1405)_b$ state does not.
Hence, $\Lambda(1405)_a$ shares the same nature of the other states of the 
unnatural parity parent Regge trajectory, {\it i.e.} it is mostly a three-quark state,
while $\Lambda(1405)_b$ does not belong to any linear Regge trajectory and, 
hence, cannot be a three-quark state, {\it i.e.} it is either a molecule or a pentaquark.
We refer the reader to \cite{JPACLambda1405} for a quantitative analysis and details on the arguments.
This result is consistent with quark-diquark models finding only one $\Lambda$(1405) state \cite{Santopinto14,Faustov15}
and different results in the exploratory lattice QCD calculations \cite{Hall2015,Engel13}.
Further studies should assess if the nature of $\Lambda (1405)_b$
is that of a pentaquark or a molecular state.

%%%%%%%%%%%%%%%%%%%%%%%%%%%%%%%%
%	Figure: Regge plots
%%%%%%%%%%%%%%%%%%%%%%%%%%%%%%%%

%%%%%%%%%%%%%%%%%%%%%%%%%%%%%%%%
%	Conclusions
%%%%%%%%%%%%%%%%%%%%%%%%%%%%%%%%
\section{Conclusions}
\begin{enumerate}
\item We have developed a $\bar{K}N$ amplitude in the resonance region that
incorporates up to 13 channels per partial wave, has the right angular momentum barrier, analyticity and unitarity,
obtaining the most comprehensive picture of the $S = -1$ hyperon spectrum to the date;
\item The model can be incorporated in the analysis of three-body decay experiments for pentaquark searches and 
 two kaon photoproduction experiments for strangeonia and exotics with hidden strangeness search;
\item The codes to compute the partial waves and the observables (cross sections and asymmetries) 
can be run online and downloaded from \cite{JPACWebpage};
\item Using Regge phenomenology and our knowledge of the $\Lambda^*$ spectrum 
we are able to establish that the higher-mass
$\Lambda$(1405) resonance is mostly a three-quark state while the lower-mass 
$\Lambda$(1405) resonance is either a molecule or a pentaquark.
\end{enumerate}

%%%%%%%%%%%%%%%%%%%%%%%%%%%%%%%%
%	Ack
%%%%%%%%%%%%%%%%%%%%%%%%%%%%%%%%
\ack
This work is part of the efforts of the Joint Physics Analysis Center.
We thank Ra\'ul Brice\~no for useful discussions.
We thank the organizers for their invitation to the conference and their warm hospitality at Cocoyoc.

%%%%%%%%%%%%%%%%%%%%%%%%%%%%%%%%
%	References
%%%%%%%%%%%%%%%%%%%%%%%%%%%%%%%%

\end{document}